\documentclass[aps,amssymb,prl,twocolumn,superscriptaddress]{revtex4}
\usepackage{graphicx}
\begin{document}
\title{Magnon Heat Transport in doped $\rm La_2CuO_4$}

\author{C. Hess}
\email[]{christian.hess@physics.unige.ch}
\affiliation{2. Physikalisches Institut, RWTH-Aachen, 52056 Aachen, Germany}

\affiliation{II. Physikalisches Institut, Universit\"{a}t zu K\"{o}ln, 50937 K\"{o}ln, Germany}

\author{B. B\"{u}chner}
\affiliation{2. Physikalisches Institut, RWTH-Aachen, 52056 Aachen, Germany}

\author{U. Ammerahl}
\affiliation{}

\author{L. Colonescu}
\affiliation{II. Physikalisches Institut, Universit\"{a}t zu K\"{o}ln, 50937 K\"{o}ln, Germany}

\author{F. Heidrich-Meisner}
\affiliation{}

\author{W. Brenig}
\affiliation{Institut f\"{u}r Theoretische Physik, Technische Universit\"{a}t Braunschweig, 38106
Braunschweig, Germany}

\author{A. Revcolevschi}
\affiliation{Laboratoire de Physico-Chimie, Universit\'{e} Paris-Sud, 91405 Orsay, France}

\date{\today}

\begin{abstract}
We present results of the thermal conductivity of $\rm La_2CuO_4$ and $\rm La_{1.8}Eu_{0.2}CuO_4$
single-crystals which represent model systems for the two-dimensional spin-1/2 Heisenberg
antiferromagnet on a square lattice. We find large anisotropies of the thermal conductivity which
are explained in terms of two-dimensional heat conduction by magnons within the CuO$_2$ planes.
Non-magnetic Zn substituted for Cu gradually suppresses this magnon thermal conductivity
$\kappa_{\mathrm{mag}}$. A semiclassical analysis of $\kappa_{\mathrm{mag}}$ is shown to yield a
magnon mean free path which scales linearly with the reciprocal concentration of Zn-ions.
\end{abstract}


\maketitle

Apart from their intimate relation to the physics of the copper-oxide high-temperature
superconductors low dimensional quantum magnets are at the center of many novel phenomena. Among
them are magnetic contributions to the thermal conductivity $\kappa$ of an unprecedentedly large
magnitude, which have been discovered recently in the quasi one-dimensional (1D) spin ladder
compound $\rm (Sr,Ca,La)_{14}Cu_{24}O_{41}$ \cite{Hess01,Sologubenko00}. Similar findings of large
and quasi 1D magnetic heat conduction have been reported also for cuprate spin-chain compounds
\cite{Sologubenko00a,Sologubenko01}. In this context it has become a key issue whether thermal
transport due to magnetic degrees of freedom in certain classes of spin models may be intrinsically
quasi-{\it ballistic}, i.e., dissipationless, thereby leading to high magnetic thermal
conductivities $\kappa_{\mathrm{mag}}$ \cite{Zotos97,Alvarez02}. Of particular interest is the mean
free path $l_{\mathrm{mag}}$ of the heat-current carrying magnetic excitations. Based on a kinetic
approach it has been shown recently \cite{Hess01}, that $l_{\mathrm{mag}}$ for the spin ladder
compound is of the order of 3000~{\AA} . Direct microscopic evaluation of the mean free path is an
open issue yet, with conflicting conclusions published \cite{Alvarez02}.


In this letter we turn to two-dimensional (2D) quantum magnets by investigating the thermal
conductivity of $\rm La_2CuO_4$ which realizes the spin-1/2 Heisenberg-antiferromagnet on the
square lattice. The magnetic structure consists of Cu$^{2+}$-ions in CuO$_2$-planes extending along
the crystallographic $a$- and $b$-directions. A strong antiferromagnetic intraplanar exchange
coupling $J/k_B\approx1550$~K \cite{Hayden91a} is present whereas the interplanar exchange
$J_\perp$ is negligible ($J_\perp /J\approx 10^{-5}$) \cite{Thio88}.

Our primary focus will be on a broad peak at high temperatures ($T$) in $\kappa(T)$ which we
observe if measured by a thermal current parallel to the CuO$_2$-planes, but which is absent for
the perpendicular direction. This is consistent with previous results obtained in Ref.
\cite{Nakamura91}. Several attempts to explain this peak have been made, e.g. scattering processes
of acoustic phonons with soft optical phonons \cite{Cohn95} or magnons \cite{Morelli89}. It has
been speculated also, that the high-$T$-maximum could be related to heat carried by magnetic
excitations \cite{Nakamura91}. However, its origin has never been elucidated unambiguously. Here,
and employing the effects of doping by various types of impurities we provide clear evidence in
favor of the anisotropic heat conduction to be due to 2D magnons indeed. We extract the mean free
path from our data using an approach similar to that for the case of $\rm
(Sr,Ca,La)_{14}Cu_{24}O_{41}$ \cite{Hess01}. It will be shown that $l_{\mathrm{mag}}$ does not only
scale linearly with the reciprocal concentration of magnetic impurities which are generated by
Zn-ions, but moreover, that it is roughly equal to the mean unidirectional distance between these
ions.


We have measured $\kappa$ of $\rm La_2CuO_4$ and $\rm La_{1.8}Eu_{0.2}CuO_{4}$ single crystals as
well as $\rm La_2Cu_{1-z}Zn_zO_4$ ($z=0$, 0.005, 0.008, 0.01, 0.02, 0.05) polycrystals as a
function of $T$. Stoichiometric oxygen contents were achieved by annealing in high vacuum
($p<10^{-4}$~mbar) for 1-3 hours at 800$^\circ$ C. The preparation of the samples \cite{Hucker02}
and the experimental method \cite{Hess01} have been described elsewhere.


\begin{figure}
\includegraphics [width=\columnwidth,clip] {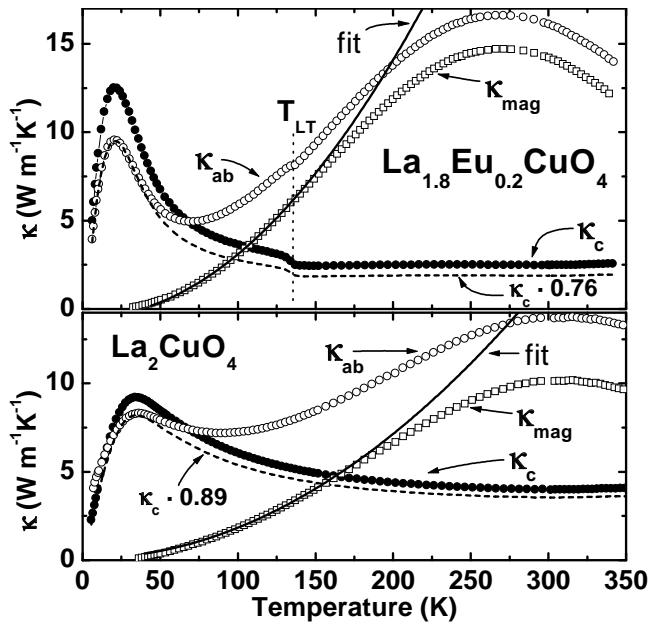}
\caption{\label{fig1}Thermal conductivity of $\rm La_2CuO_4$ (bottom) and $\rm
La_{1.8}Eu_{0.2}CuO_4$ (top). Full circles: $\kappa_c$. Open circles: $\kappa_{ab}$.  Open
squares: $\kappa_{\mathrm{mag}}$. Solid line: fit according to Eq.~\ref{fit2d}. Dashed line:
$\kappa_{ab,\mathrm {ph}}$.}
\end{figure}
In the lower panel of Fig.~\ref{fig1} we present the thermal conductivity of single crystalline
$\rm La_2CuO_4$ if measured parallel ($\kappa_{ab}$) and perpendicular ($\kappa_c$) to the
CuO$_2$-planes as a function of $T$. The pronounced anisotropy of $\kappa$ strongly resembles that
of spin ladder compounds \cite{Hess01}: $\kappa_c$ exhibits a $T$-dependence which is typical for
insulating crystalline materials with pure phononic thermal conductivity $\kappa_{\mathrm {ph}}$.
In contrast to this, the low-$T$ peak of $\kappa_{ab}$ is followed by a minimum around 80~K and a
strong increase of $\kappa_{ab}$ which eventually develops into a broad peak around 300~K that
clearly exceeds the low-$T$ peak.

Several conventional mechanisms which may be invoked to explain this unusual $T$-dependence and the
resulting anisotropy can be dismissed. First, we can exclude effects of radiative heat transport
since the optical properties of $\rm La_2CuO_4$ are almost isotropic in the relevant energy-range
below $\rm h\nu\lesssim0.1$~eV \cite{Uchida91}. Second, electronic contributions to the thermal
current can be excluded since the material is insulating. Therefore, phononic heat transport is
expected to dominate. However, and third, a scenario solely based on thermal conduction by acoustic
phonons is very unlikely since an increase of scattering usually causes a negative slope of
$\kappa(T)$ for intermediate and higher $T$, in contrast to our observation. Hence, either an
unusual scattering process acts on the acoustic phonons or an additional contribution apart from
the conventional phononic background must be present for $\kappa_{ab}$.

We now turn to the upper panel of Fig.~\ref{fig1} where $\kappa_{ab}(T)$ and $\kappa_c(T)$ of $\rm
La_{1.8}Eu_{0.2}CuO_{4}$ are shown. Both curves strongly resemble our findings for
$\rm La_2CuO_4$, yet exhibiting obvious differences: the low $T$-peaks of $\kappa_{ab}$ and
$\kappa_c$ are slightly
larger and more sharply shaped than in the case of $\rm La_2CuO_4$. Furthermore, a step-like anomaly
is present at $T_{LT}\approx135$~K. Above $T_{LT}$ $\kappa_c$ remains almost constant and stays
below the value for the undoped case, while $\kappa_{ab}$ also exhibits a high-$T$ maximum at
$T\approx270$~K which is even larger than in $\rm La_2CuO_4$.

The difference between $\kappa_c$ of Eu-doped and of pure $\rm La_2CuO_4$ can be attributed to a
difference in phononic heat conduction. Upon doping $\rm La_2CuO_4$ with Eu, enhanced scattering of
phonons reduces $\kappa_c$ for $T>T_{LT}$, where both compounds have the same structure. The
anomaly at $T_{LT}$ ($\rm La_{1.8}Eu_{0.2}CuO_{4}$) signals the transition to a new structural
phase for $T<T_{LT}$ where $\kappa_{\mathrm {ph}}$ is enhanced. The structural peculiarities of
rare earth doped cuprates are well known \cite{Buchner94}. Their influence on $\kappa_{\mathrm
{ph}}$ will be discussed in detail in a forthcoming paper \cite{Hess03}.

Despite the increase of scattering of phonons due to Eu-impurities in $\rm La_{1.8}Eu_{0.2}CuO_{4}$
the high-temperature peak of $\kappa_{ab}$ is of larger magnitude than in $\rm La_2CuO_4$. This
excludes the peak to stem from the previously mentioned heat transport by acoustic phonons. The
high-$T$ peaks in $\kappa_{ab}$ of both compounds therefore must originate from a different heat
transport channel. In principle, two different kinds of excitation could generate such an
additional heat current. One possibility is heat transport by dispersive optical phonons, the other one is thermal current
carried by magnetic excitations. Magnetic heat transport is the most likely explanation, since the magnetism is truly two-dimensional and the
refore could explain the observed anisotropy. This is not true for dispersive optical phonons, which certainly are present in this material, but are found along all crystal axes \cite{Pintschovius91}.

This result is corroborated by the strong suppression of the high-$T$ peak in hole-doped $\rm La_{2-x}Sr_xCuO_{4}$ \cite{Nakamura91}, even at a very low Sr-content of $x\approx0.01$ \cite{Hess03}. On the one hand such an effect of
doping is not to be expected in the case of a high-$T$ peak originating
from optical phonons, since the lattice impurities induced by the Sr-ions are similar to the Eu-impurities
discussed above. On the other hand the strong frustration of antiferromagnetism upon doping of mobile holes
(cf. Ref.~\cite{Hucker02} and references therein) would provide for a satisfying explanation for
the suppression of a peak of magnetic origin. Note in this context that Eu-doping leaves the
CuO$_2$-planes and therefore the magnetism almost unaffected. Hence we conclude, that the high-$T$
peak of $\kappa_{ab}$ originates from magnetic excitations which propagate only within the CuO$_2$-planes.  
$\kappa_{ab}$ therefore consists of a usual phonon background and a magnon
contribution $\kappa_{\mathrm{mag}}$ while $\kappa_c$ is purely phononic.

Upon applying a magnetic field of 8~T no significant changes of $\kappa$ were detected. This
is, however, consistent with a magnetic origin of the high-$T$ peak since the corresponding
Zeeman-energy is orders of magnitude smaller than the magnetic exchange coupling $J$.

In order to extract $\kappa_{\mathrm{mag}}$ from $\kappa_{ab}$ we make use of the anisotropy of
$\kappa$ assuming that the phononic part $\kappa_{ab,\mathrm {ph}}$ of $\kappa_{ab}$ is roughly
proportional to $\kappa_c$ which is justified by only weakly
anisotropic elastic constants \cite{Pintschovius91}. Since the magnetic contributions roughly follow a $T^2$-law (see
below) they are expected to be negligible in the range of the low-$T$ peak. Therefore, for $\rm
La_2CuO_4$ as well as for $\rm La_{1.8}Eu_{0.2}CuO_{4}$ a reasonable estimate of
$\kappa_{ab,\mathrm {ph}}$ is achieved by scaling the corresponding data for $\kappa_c$ such as to
match its low-$T$ peaks with that of $\kappa_{ab}$. In Fig. \ref{fig1} the data thus obtained for
$\kappa_{ab,\mathrm {ph}}$ of both compounds are represented by dashed lines.
$\kappa_{\mathrm{mag}}$ was extracted by subtracting $\kappa_{ab,\mathrm {ph}}$ from $\kappa_{ab}$
(open squares in Fig. \ref{fig1}). This procedure involves significant uncertainties only at low
$T$ where $\kappa_{\mathrm{mag}}\ll\kappa_{ab,\mathrm {ph}}$. At intermediate and higher $T$ the
relative uncertainties are smaller and rather concern the magnitude of $\kappa_{ab,\mathrm {ph}}$
than its slope. Therefore we assume that in this $T$-range the difference between the extracted and
the true $\kappa_{\mathrm{mag}}$ is roughly constant.

\begin{table}
\center{\begin{tabular}{lll}\hline\hline \multicolumn{1}{c}{Sample}&\multicolumn{1}{c}{$l_{\mathrm{mag}}$
(\AA)}&\multicolumn{1}{c}{Fit interval (K)}\\
\hline S $\rm La_2CuO_4$&$558\pm 140$&\multicolumn{1}{c}{70--158}\\
S $\rm La_{1.8}Eu_{0.2}CuO_{4}$&$1157\pm60$&\multicolumn{1}{c}{54--131}\\
\hline P $\rm La_2CuO_4$ &\multicolumn{1}{c}{815}&\multicolumn{1}{c}{70--140}\\
P $\rm La_2Cu_{0.995}Zn_{0.005}O_4$&\multicolumn{1}{c}{515}&\multicolumn{1}{c}{70--158}\\
P $\rm La_2Cu_{0.992}Zn_{0.008}O_4$&\multicolumn{1}{c}{435}&\multicolumn{1}{c}{82--164}\\
P $\rm La_2Cu_{0.99}Zn_{0.01}O_4$&\multicolumn{1}{c}{266}&\multicolumn{1}{c}{100--173}\\
P $\rm La_2Cu_{0.98}Zn_{0.02}O_4$&\multicolumn{1}{c}{153}&\multicolumn{1}{c}{118--200}\\
P $\rm La_2Cu_{0.95}Zn_{0.05}O_4$&\multicolumn{1}{c}{49}&\multicolumn{1}{c}{141--223}\\
\hline\hline
\end{tabular}}
\caption{\label{tab1} Fit parameter of $\kappa_{\mathrm{mag}}$ for single (S) and polycrystals (P). The relative errors for $l_{\mathrm{mag}}$ of the polycrystals amount to about
10\%. In all cases the errors 
have been obtained using different estimates for $\kappa_{\mathrm {ph}}$.}
\end{table}


Following standard linearized Boltzmann-theory we have $\varkappa^i=\frac{1}{2}\frac{1}{(2\pi)^2}\int
v_{\bf k}l_{\bf k}\epsilon_{\bf k} \frac{d}{dT}n_{\bf k}d{\bf k}$ for the 2D thermal conductivity of a single magnon
dispersion branch (labelled by $i$), where $v_{\bf k}$, $l_{\bf k}$, $\epsilon_{\bf k}$ and $n_{\bf k}$ denote velocity, mean free path, energy and Bose-function of a
magnon. Note that $\kappa_{\mathrm{mag}}^i$ of a three-dimensional ensemble of
planes, as realized in $\rm La_2CuO_4$, is given by $\kappa_{\mathrm{mag}}^i=\frac{2}{c}\varkappa^i$, where
$c=13.2$~{\AA} is the lattice constant of $\rm La_2CuO_4$ perpendicular to the planes. Then the
total $\kappa_{\mathrm{mag}}$ results from summing up $\kappa_{\mathrm{mag}}^i$ of each magnon
branch.

%

In order to calculate $\kappa_{\mathrm{mag}}^i$ we approximate the magnon dispersion relation
$\epsilon_{\bf k}$ of the two branches $i=1,2$ with the 2D-isotropic expression $\epsilon_{\bf k}=\epsilon_k=\sqrt{\Delta^2_i+(\hbar v_0k)^2}$, which describes the dispersion observed experimentally \cite{Keimer93,Coldea01} for
small values of $k$. Here, $v_0$ is the spin wave velocity while $\Delta_1$ and $\Delta_2$ denote
the spin gaps of each magnon branch. For clarity, we  define the characteristic temperature
$\Theta_M=(\hbar v_0 \sqrt{\pi}) /(a k_B)$ where $a=3.8$~{\AA} is the lattice constant of the
CuO$_2$-planes \cite{note1}. Assuming a momentum independent mean free path, i.e. $l_{\bf k}\equiv
l_{\mathrm{mag}}$, we obtain for each magnon branch:
\begin{equation}
\label{fit2d}{\kappa_{\mathrm{mag}}^i=\frac{v_0k_B \,l_{\mathrm{mag}}}{2a^2c}
\frac{T^2}{\Theta_M^2}
\int\limits_{x_{\mathrm{0,i}}}^{x_{\mathrm{max}}}x^2\sqrt{x^2-x_{\mathrm{0,i}}^2}
\frac{dn(x)}{dx}dx.}
\end{equation}
Here, the integral is dimensionless but temperature dependent via $x_{\mathrm{0},i}=\Delta_i/(k_B
T)$ and $x_{\mathrm{max}}$. The upper boundary $x_{\mathrm{max}}$, however, may be set to infinity
without affecting the fit at temperatures $T\lesssim 300$~K. Since from neutron scattering
experiments $v_0\approx1.287\cdot10^5$~m/s \cite{Hayden91a} as well as $\Delta_{1}/k_B\approx26$~K
and $\Delta_{2}/k_B\approx58$~K \cite{Keimer93} are well known quantities, $l_{\mathrm{mag}}$ is
the only unknown parameter in Eq.\ \ref{fit2d}. Assuming $l_{\mathrm{mag}}$ to be $T$-independent
at low $T$, we can use Eq.\ \ref{fit2d} in order to check the $T$-dependence of our experimental
$\kappa_{\mathrm{mag}}$, which for $T\gtrsim\Delta_{1,2}$ should roughly be
$\kappa_{\mathrm{mag}}\propto T^2$, as the integral is only weakly $T$-dependent in this
$T$-range. Furthermore, $l_{\mathrm{mag}}$ can be extracted  from the data.
For $T\gtrsim250$~K interactions between the magnons become important \cite{Keimer92} which
requires a renormalization of the spin wave parameters and therefore we restrict the application
of Eq. \ref{fit2d} with constant $l_{\mathrm{mag}}$ to $T\lesssim250$~K.


Fitting Eq.\ (\ref{fit2d}) to the experimental data we allow for an additive shift of the
$\kappa_{\mathrm{mag}}$-curve which yields a further free parameter apart from $l_{\mathrm{mag}}$
and accounts for the aforementioned uncertainties in the magnitude of $\kappa_{\mathrm{mag}}$ \cite{note2}. For
both compounds satisfactory fits were obtained at intermediate $T$-ranges. These are
listed in Table \ref{tab1}, together with the resulting values for $l_{\mathrm{mag}}$. The fitting
curves are represented by solid lines in Fig.~\ref{fig1}. While the slight deviations between the
fitted and experimental data towards low $T$ are due to the uncertainties in $\kappa_{ab,\mathrm
{ph}}$ in this range, the deviations at high $T$ can be understood in terms of the $T$-dependence of $l_{\mathrm{mag}}$ due to enhanced magnon-magnon scattering.

\begin{figure}
\includegraphics [width=\columnwidth,clip] {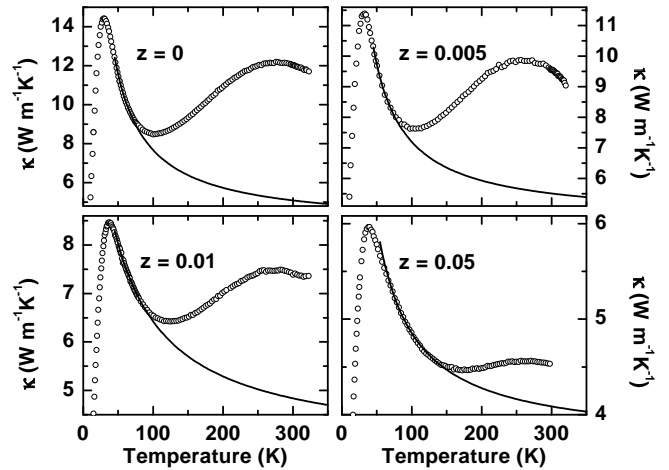}
\caption{\label{fig2}Open circles: Thermal conductivity $\kappa$ of $\rm La_2Cu_{1-z}Zn_zO_4$
polycrystals ($z=0$, 0.005, 0.01, 0.05) as a function of $T$. Solid lines: extrapolated
$\kappa_{\mathrm {ph}}$.}
\end{figure}

The data are consistent with a constant $l_{\mathrm{mag}}$ for $T$ both, within the fit interval
and below, indicating that in this range $T$-dependent scattering processes like magnon-magnon
scattering or even magnon-phonon scattering may be discarded. Therefore, relevant processes seem
to be sample-boundary scattering or scattering at static magnetic defects. For
$\rm La_{1.8}Eu_{0.2}CuO_{4}$
and $\rm La_2CuO_4$ we find $l_{\mathrm{mag}}\approx1160$~{\AA} and
$l_{\mathrm{mag}}\approx560$~{\AA}, respectively. Since these values are far too small to
correspond to the crystal dimensions, magnon-defect scattering is the most likely candidate.
This conclusion is consistent with the fact that $\kappa_{\mathrm{mag}}$ and
$l_{\mathrm{mag}}$ are quantitatively different for $\rm La_{1.8}Eu_{0.2}CuO_{4}$ and
$\rm La_2CuO_4$: since the
magnetic properties of both compounds are expected to be identical in essence, unequal
$\kappa_{\mathrm{mag}}$ can arise only due to a difference in densities of the
magnetic defects that restrict $l_{\mathrm{mag}}$.

In order to check our analysis quantitatively it would be desirable to measure the density of
static magnetic defects of our crystals. In lack of such methods we have performed measurements of
$\kappa$ on samples with a well defined density of magnetic defects. Such defects can be induced in
$\rm La_2CuO_4$ by substituting a small amount of Cu$^{2+}$-ions by non-magnetic Zn$^{2+}$-ions.
Representative results on our polycrystalline samples of $\rm La_2Cu_{1-z}Zn_zO_4$ are presented in
Fig. \ref{fig2}. As is expected for doping of static structural and magnetic defects, the Zn-impurities lead to a
gradual suppression of both, the phonon as well as the magnon contribution to $\kappa$ \cite{notesrzn}. Due to the
polycrystalline nature of our samples anisotropic information on $\kappa$ is averaged over. We
therefore estimate the phonon contributions $\kappa_{\mathrm {ph}}^{\mathrm {poly}}$ by fitting
$\kappa$ right of its maximum by $\kappa_{\mathrm {ph}}=\alpha/T+\beta$ and by extrapolating this
fit towards high $T$ (solid lines in Fig.\ \ref{fig2}). In turn, $\kappa_{\mathrm{mag}}^{\mathrm
{poly}}$ on the polycrystals is obtained by $\kappa_{\mathrm{mag}}^{\mathrm
{poly}}=\kappa-\kappa_{\mathrm {ph}}$.

\begin{figure}
\includegraphics [width=\columnwidth,clip] {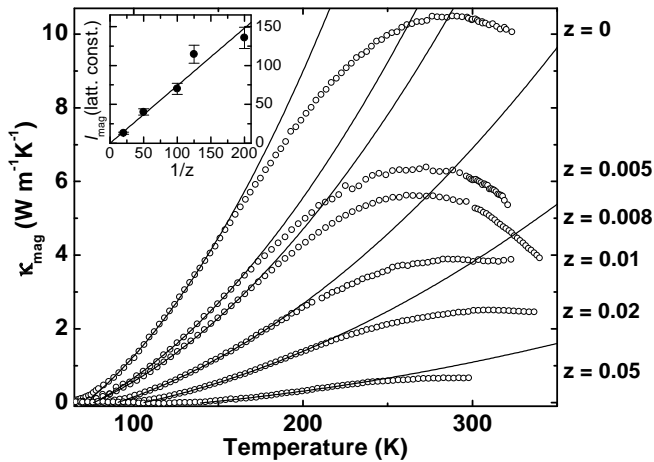}
\caption{\label{fig3}Main panel: Magnon thermal conductivity $\kappa_{\mathrm{mag}}$ of $\rm
La_2Cu_{1-z}Zn_zO_4$ ($z=0$, 0.005, 0.008, 0.01, 0.02, 0.05) as a function of $T$ (open circles).
Solid lines: fits according to Eq.~\ref{fit2d}. Inset: $l_{\mathrm{mag}}$ as a function of $1/z$ in
units of lattice constants $a$. Solid line: fit line through origin.}
\end{figure}
Note, that the measured $\kappa_{\mathrm{mag}}^{\mathrm {poly}}$ is smaller than the intrinsic
$\kappa_{\mathrm{mag}}$ of these compounds by the factor of 2/3 due to averaging over all three
components of the $\kappa$ tensor \cite{note3}. As shown in Fig.\ \ref{fig3},
$\kappa_{\mathrm{mag}}=(3/2)\kappa_{\mathrm{mag}}^{\mathrm {poly}}$ systematically decreases with
increasing Zn-content as it is expected for static defects. Analyzing these data we proceed
analogous to the undoped case by using Eq. \ref{fit2d}. However, we neglect a slight reduction of
the spin wave velocity \cite{Brenig91} and changes of the spin gaps induced by the Zn-ions. These
effects lead to corrections smaller than the experimental error. The fits are represented by the
solid lines in Fig. \ref{fig3}, the resulting $l_{\mathrm{mag}}$ and the fit intervals are
reproduced in Table \ref{tab1}. For typical transport experiments it is expected that the mean free
path of the quasi-particles is proportional to the reciprocal defect concentration, i.e., to $1/z$.
We therefore plot $l_{\mathrm{mag}}$ as a function of $1/z$ in the inset of Fig. \ref{fig3}. The
linear scaling between $l_{\mathrm{mag}}$ and $1/z$ is clearly confirmed. From a linear fit
including the origin (solid line) we find that $l_{\mathrm{mag}}\approx 0.74 a/z$. Since $1/z$ is
equal to the mean unidirectional distance of Zn-ions, i.e. static magnetic defects,
$l_{\mathrm{mag}}$ gives a direct measure of these distances \cite{note4}. Therefore, this result
confirms the above quantitative analysis of $\kappa_{\mathrm{mag}}$ of the single crystals based on
Eq. \ref{fit2d}.
In conclusion, both, qualitatively as well as quantitatively our results strongly suggest, that
the high-$T$ peak observed in $\kappa_{ab}$ of doped and undoped $\rm La_2CuO_4$ arises due to
magnon heat transport which is confined to the CuO$_2$-planes. 
 We note, that in general a large $\kappa_{\mathrm{mag}}$ is rarely observed at high $T$ and requires a unique synergy of magnon-phonon coupling \cite{Sanders77}, large $l_{\mathrm{mag}}$ and high spin wave velocity $v_0$. Its realization in $\rm La_2CuO_4$ may lead to future use of magnetic heat transport as a
tool to study the interactions of magnetic excitations with other quasiparticles like
holes or phonons.

We acknowledge support by the DFG through SP1073.

\bibliographystyle{apsrev}

\begin{thebibliography}{25}
\bibitem{note1}{We neglect the slight orthorhombicity of $\rm
La_2CuO_4$.}
\bibitem{note3}{$\kappa^{\mathrm
{poly}}=(2\kappa_{ab}+\kappa_c)/3=(2\kappa_{\mathrm{mag}}+2\kappa_{ab,\mathrm {ph}}+\kappa_c)/3$.
Since $\kappa_{\mathrm {ph}}^{\mathrm {poly}}=(2\kappa_{ab,\mathrm {ph}}+\kappa_c)/3$ we find
$\kappa_{\mathrm{mag}}=(3/2)\kappa_{\mathrm{mag}}^{\mathrm {poly}}$.}
\bibitem{note4}{The extracted
proportional constant might be reduced slightly due to the polycrystallinity of our samples.}
\bibitem{notesrzn}{It is natural to ascribe the different dependence of $\kappa_{\mathrm {mag}}$ on Zn- and Sr-doping to the correspondingly induced static and mobile magnetic defects.}
\end{thebibliography}
\end{document}